\documentstyle[floats,prl,aps,epsf,preprint]{revtex}
\begin{document}
\bibliographystyle{plain}
\title{Remark on the effective potential of the gravitational perturbation 
in the black hole background projected on the brane} 
\author{ 
D. K. Park\footnote{Email:dkpark@hep.kyungnam.ac.kr 
}}
\address{Department of Physics, Kyungnam University,
Masan, 631-701, Korea.}
\date{\today}
\maketitle

\begin{abstract}
The polar perturbation is examined when the spacetime is expressed by a 
$4d$ metric induced from higher-dimensional Schwarzschild geometry.
Since the spacetime background is not a vacuum solution of $4d$
Einstein equation, the various general principles are used to understand
the behavior of the energy-momentum tensor under the perturbation. It is 
found that although the general principles fix many components, they
cannot fix two components of the energy-momentum tensor. Choosing two 
components suitably, we derive the effective potential which has a 
correct $4d$ limit.
\end{abstract}

\newpage
The spacetime stability/instability problem is an important issue to determine
the final state of the gravitational collapse. It is also important in the 
astrophysical side for the detection of the gravitational wave. Long 
ago this issue was firstly examined by Regge and Wheeler\cite{reg57} in
the background of the $4d$ Schwarzschild geometry. In four spacetime dimension
there are two types of the gravitational perturbations called 
axial (or odd-parity) and polar (or even-parity) perturbations\cite{chand83}.
The Schwarzschild geometry in general admits the perturbed equations to 
be separable and the resultant radial equations can be transformed into the
Schr\"{o}dinger-like form. Thus the effective potential for each perturbation
was explicitly derived in Ref.\cite{vish70,zer70}. The transformation into
the Schr\"{o}dinger-like expression is important to predict the various
physical phenomena because we have much background in quantum mechanics.

Recent quantum gravity such as string theories\cite{polchin98} and the 
brane-world scenarios\cite{ark98-1,rs99-1} in general introduce the 
extra dimensions to reconcile general relativity with quantum physics.
In this context, recently, much attention is paid to the higher-dimensional
spacetime. The various spacetime metrics of the higher-dimensional 
black holes were derived in Ref.\cite{tang63,myers86}. The gravitational 
perturbations in the higher-dimensional Schwarzschild background were also
discussed in Ref.\cite{koda03}. When the spacetime dimension is more than
four, there are three types of the gravitational perturbations called
scalar, vector, and tensor perturbations. The scalar and vector perturbations 
correspond to the polar and the axial perturbations in four dimension. There
is no correspondence of the tensor perturbation in four dimension.

Recently, the axial perturbation is discussed when the spacetime is a 
$4d$ metric induced from the higher-dimensional Schwarzschild 
black hole\cite{park05-1}. In the following we will briefly review 
Ref.\cite{park05-1}. 
Subsequently the polar perturbation will be addressed
in this background.

We start with an $4d$ metric which is induced from the $(4+n)$-dimensional 
Schwarzschild black hole
\begin{equation}
\label{metric1}
ds^2 = -h(r) dt^2 + h^{-1}(r) dr^2 + r^2 (d \theta^2 + \sin^2 \theta d\phi^2)
\end{equation}
where $h(r) = 1 - (r_H / r)^{n+1}$. As mentioned, 
there are two types of the gravitational metric perturbations 
in four spacetime dimension called 
axial and polar perturbations\cite{reg57,chand83}. In the former case the metric
is changed into $ds^2 + \delta s_A^2$ where
\begin{equation}
\label{metric2}
\delta s_A^2 = \left[ H_0(r) dt d\phi + H_1(r) dr d\phi \right] e^{i \omega t}
\sin \theta \frac{d P_{\ell}}{d \theta} (\cos \theta)
\end{equation}
while the metric for the latter case is $ds^2 + \delta s_P^2$ where 
\begin{equation}
\label{metric3}
\delta s_P^2 = \left[ H(r) h dt^2 + H(r) h^{-1} dr^2 + r^2 K(r)
(d \theta^2 + \sin^2 \theta d\phi^2) + 2 H_1(r) dt dr \right] e^{i \omega t}
P_{\ell} (\cos \theta)
\end{equation}
where $P_{\ell} (\cos \theta)$ is an usual Legendre polynomial. The various 
$r$-dependent functions are assumed to be small for the linearization.  

When $n=0$, the metric in Eq.(\ref{metric1}) is an usual $4d$ Schwarzschild
metric which is a vacuum solution of the Einstein equation. In this $4d$ case,
therefore, the gravitational linearized fluctuation can be expressed as 
$\delta R_{\mu \nu} = 0$ where $R_{\mu \nu}$ is a Ricci tensor. Using this 
representation the effective potentials for the perturbations were derived
long ago\cite{vish70,zer70}
\begin{eqnarray}
\label{potential1}
V_A(r)&=& h \left[\frac{2 \lambda + 2}{r^2} - \frac{3}{r^2} 
                                 \left(\frac{r_H}{r}\right) \right]
                                                 \\   \nonumber
V_P(r)&=& \frac{h}{(2 \lambda r + 3 r_H)^2}
\left[8 \lambda^2 (\lambda + 1) + 12 + 12 \lambda^2 \left(\frac{r_H}{r}\right)
      + 18 \lambda \left(\frac{r_H}{r}\right)^2 + 
        9 \left(\frac{r_H}{r}\right)^3 \right]
\end{eqnarray}
where $\lambda = (\ell - 1) (\ell + 2) / 2$. 

The most interesting feature of the potentials is that they are related to 
each other as following\cite{chand83}
\begin{equation}
\label{potential2}
V_{P,A} (r) = \pm \beta \frac{d f}{d r_*} + \beta^2 f^2 + \kappa f
\end{equation}
where the upper(lower) sign corresponds to the polar(axial) 
perturbation and $r_*$ is a
``tortoise'' coordinate defined $d r / d r_* = h$. In Eq.(\ref{potential2})
$\beta = 3 r_H$, $\kappa = 4 \lambda (\lambda + 1)$ and 
\begin{equation}
\label{definef}
f = \frac{h}{r (2 \lambda r + 3 r_H)}.
\end{equation}
In fact this relation was found when the Newman-Penrose 
formalism\cite{newman62} is applied to the gravitational perturbations. This
explicit relation between $V_P$ and $V_A$ enables us to realize that both
potentials have the same transmission coefficient\footnote{In
Ref.\cite{chand83} the general criterion for the same transmission 
coefficient of the different one-dimensional potentials was derived in 
terms of the KdV equation, which is well-known in the solitonic theories.}.

In this context the following several questions arise naturally: 
(i) Do the effective potentials exist when the spacetime is a metric induced 
from the higher-dimensional Schwarzschild black hole?
(ii) If the potentials exist, is there any relation between them like
     Eq.(\ref{potential2})?
(iii) Do they have same transmission coefficient like $4d$ case?  
We would like to address the first issue as much as possible in this letter.

The most difficult problem one copes with when the spacetime is induced 
from the higher-dimensional one is the fact that Eq.(\ref{metric1}) is not 
a vacuum solution of the $4d$ Einstein equation if $n \neq 0$. Thus what 
we can do is to assume that Eq.(\ref{metric1}) is a non-vacuum solution,
{\it i.e.} ${\cal E}_{\mu \nu} = T_{\mu \nu}$ where
${\cal E}_{\mu \nu}$ and $T_{\mu \nu}$ are Einstein and energy-momentum
tensors, respectively. Since, however, $T_{\mu \nu}$ is not 
originated from the real matter living on the brane, but appears
effectively in the course of the projection of the bulk metric to the 
brane, we do not know how $T_{\mu \nu}$ is transformed when the perturbations 
(\ref{metric2}) or (\ref{metric3}) is turned on. The only way, in our opinion,
to get rid of this difficulty is to get an information on the energy-momentum
tensor as much as possible from the general principles.

The fluctuation equation in the background of the non-vacuum solution is 
generally expressed as $\delta {\cal E}_{\mu \nu} = \delta T_{\mu \nu}$.
Thus the difficulty mentioned above is how to obtain $\delta T_{\mu \nu}$
from the general principles without knowing the exact nature of the matter.

The axial perturbation in the background of Eq.(\ref{metric1}) was 
discussed in Ref.\cite{park05-1}. In this case the non-vanishing components
of $\delta T_{\mu \nu}$ are $\delta T_{t \phi}$, $\delta T_{r \phi}$ and 
$\delta T_{\theta \phi}$. The general principles used in
Ref.\cite{park05-1}: (i) covariant conservation of $T^{\mu \nu}$
(ii) linear dependence of the Einstein equation (iii) the correct $4d$ limit
of the effective potential. It turned out that the first principle generates
the unique non-trivial constraint to the energy-momentum tensor. 
But the second one
is automatically satisfied if the first principle is used. Third one makes
a constraint only in the $4d$ limit of the energy-momentum tensor. Thus the
principle used in Ref.\cite{park05-1} could not completely determine 
$\delta T_{t \phi}$, $\delta T_{r \phi}$ and 
$\delta T_{\theta \phi}$. Thus what we can do in this situation is to choose 
the energy-momentum tensor consistently with the general principles. In 
Ref.\cite{park05-1} $\delta T_{r \phi} = \delta T_{\theta \phi} = 0$ is chosen
for simplicity. Then the effective potential for the axial perturbation 
becomes 
\begin{equation}
\label{potential3}
V_A(r) = h(r) \left[ \frac{2 \lambda + 2}{r^2} - 
           \left\{(n+1)^2 + 2\right\} \left( \frac{r_H}{r} \right)^{n+1}
                                                  \right].
\end{equation}
It is easy to show that Eq.(\ref{potential3}) has a correct $4d$ limit. 
But the problem is that the different choice may yield a different 
potential.

Now we would like to consider the polar perturbation by adding the perturbation 
(\ref{metric3}) into (\ref{metric1}). From the non-vanishing components of 
$\delta {\cal E}_{\mu \nu}$ in the gravity side it is reasonable to assume 
that the non-vanishing components of $\delta T_{\mu \nu}$ are $\delta T_{t t}$,
$\delta T_{t r}$, $\delta T_{t \theta}$, $\delta T_{r r}$, 
$\delta T_{r \theta}$, $\delta T_{\theta \theta}$ and $\delta T_{\phi \phi}$.
The spherical symmetry of course guarantees 
$\delta T_{\phi \phi} = \delta T_{\theta \theta} \sin^2 \theta$.

The $(t, r)$, $(t, \theta)$ and $(r, \theta)$ componets of the Einstein
equations reduce to
\begin{eqnarray}
\label{first-order}
& &K' = \frac{i}{\omega} \delta \Theta_{t r} + 
   \left( \frac{h'}{2 h} - \frac{1}{r} \right) K - 
   \frac{i}{\omega} \left( \frac{h'}{r} + \frac{h + \lambda}{r^2} \right)
   H_1 + \frac{1}{r} H
                                               \\   \nonumber
& &H_1' = \frac{2}{h} \delta \Theta_{t \theta} - \frac{h'}{h} H_1 + 
\frac{i \omega}{h} (H + K)
                                               \\   \nonumber
& &H' = \frac{i}{\omega} \delta \Theta_{t r} + 2 \delta \Theta_{r \theta}
   + \left( \frac{1}{r} - \frac{h'}{h} \right) H + 
    \left( \frac{h'}{2 h} - \frac{1}{r} \right) K - \frac{i}{\omega}
   \left( \frac{h'}{r} + \frac{h + \lambda}{r^2} - \frac{\omega^2}{h}
                                                        \right) H_1
\end{eqnarray}  
where the prime denotes a differentiation with respect to $r$ and 
\begin{eqnarray}
\label{defT1}
& &\delta T_{t r} = \delta \Theta_{t r} (r) e^{i \omega t} 
                           P_{\ell} (\cos \theta)
                                                     \\  \nonumber
& &\delta T_{t \theta} = \delta \Theta_{t \theta} (r) e^{i \omega t}
                 \frac{d P_{\ell}}{d \theta} (\cos \theta)
                                                     \\  \nonumber
& &\delta T_{r \theta} = \delta \Theta_{r \theta} (r) e^{i \omega t}
                 \frac{d P_{\ell}}{d \theta} (\cos \theta).
\end{eqnarray}

Now we consider the covariant conservation of the energy-momentum tensor,
{\it i.e.} $T^{\mu \nu}_{\hspace{0.3cm}; \mu} = 0$. This gives three conditions when 
$\nu = t$, $r$, and $\theta$. Using the conditions derived when
$\nu = t$ and $\theta$, one can express $\delta T_{t t}$ and 
$\delta T_{\theta \theta}$ in terms of the non-diagonal components as 
following:
\begin{eqnarray}
\label{defT2}
\delta T_{t t}&=& -\frac{i}{\omega}
      \Bigg[h^2 \left\{ \partial_r \delta \Theta_{t r} + 
                        \left(\frac{2}{r} + \frac{h'}{h} \right) 
                                         \delta \Theta_{t r} \right\}
            -\frac{2 h}{r} \left(\frac{h + \lambda}{r} + h' \right)
                                               \delta \Theta_{t \theta}
                                                            \\   \nonumber
& & \hspace{4.5cm}
-\frac{h^2}{r} \left(h' + \frac{r}{2} h''\right)
 \left( \frac{i \omega}{h} K + \frac{2}{r} H_1 \right)
                                                       \Bigg]
e^{i \omega t} P_{\ell} (\cos \theta )
                                                           \\  \nonumber
\delta T_{\theta \theta}&=& \left[
\frac{i \omega r^2}{h} \delta \Theta_{t \theta} - r^2 h
      \left\{ \partial_r \delta \Theta_{r \theta} + 
      \left( \frac{h'}{h} + \frac{2}{r} \right) \delta \Theta_{r \theta}
                                                               \right\}
        + r \left( h' + \frac{r}{2} h'' \right) K \right] e^{i \omega t}
        P_{\ell} (\cos \theta ).
\end{eqnarray}
With an aid of Eq.(\ref{defT2}) the condition at $\nu = r$ gives the 
following differential equation for $\delta T_{r r}$:
\begin{equation}
\label{defT3}
\partial_r \delta T_{r r} + 
\left(\frac{3 h'}{2 h} + \frac{2}{r} \right) \delta T_{r r}
= \left(Q_1 (r) + Q_2 (r) \right) e^{i \omega t} P_{\ell} (\cos \theta)
\end{equation}
where
\begin{eqnarray}
\label{defQ}
Q_1(r)&=& \frac{i}{\omega} \frac{h'}{2 h}
\left[ \partial_r \delta \Theta_{t r} + 
      \left( \frac{4}{r} + \frac{h'}{h} + \frac{h''}{h'} + 
             \frac{2 \omega^2}{h h'} \right) \delta \Theta_{t r} \right]
                                                             \\  \nonumber
& &- \frac{2}{r} \left[\partial_r \delta \Theta_{r \theta} + 
                       \frac{h - \lambda}{r h} \delta \Theta_{r \theta}
                                                               \right]
- \frac{i}{\omega} \frac{1}{r h^2}
  \left[ \frac{(h + \lambda) h'}{r} + h'^2 - 2 \omega^2 \right] 
                                              \delta \Theta_{t \theta}
                                                               \\  \nonumber
Q_2(r)&=& \left(h' + \frac{r}{2} h''\right)
          \left[ \frac{3}{r^2 h} H - \frac{i}{\omega r h}
                \left(\frac{2 h'}{r} + \frac{h + \lambda}{r^2}\right) H_1
                + \frac{1}{r h} \left( \frac{h'}{h} - \frac{1}{r} \right) K
                                                           \right].
\end{eqnarray}
Thus $\delta T_{r r}$ also can be obtained from the non-diagonal components
of the energy-momentum tensor as following:
\begin{equation}
\label{bozo11}
\delta T_{r r} = \frac{1}{r^2 h^{3/2}} e^{i \omega t}
P_{\ell} (\cos \theta) \int dr (Q_1 + Q_2) r^2 h^{3/2}.
\end{equation}
Therefore all diagonal components can be calculated if $\delta \Theta_{t r}$,
$\delta \Theta_{t \theta}$ and $\delta \Theta_{r \theta}$ are known.

Now, we would like to consider the diagonal components of the 
Einstein equation. After some algebra one can show straightforwardly that the 
first-order differential equations in Eq.(\ref{first-order}) automatically
solve the $(t,t)$, $(\theta, \theta)$ and $(\phi, \phi)$ components of the
Einstein equation. The $(r, r)$ component gives a constraint to 
$\delta T_{r r}$ in a form:
\begin{equation}
\label{defT4}
\delta T_{r r} = \delta T^{(1)}_{r r} + \delta T^{(2)}_{r r}
\end{equation}
where
\begin{eqnarray}
\label{defT5}
\delta T^{(1)}_{r r}&=& \left(\frac{i h'}{2 \omega h} \delta \Theta_{t r}
                 - \frac{2}{r} \delta \Theta_{r \theta} \right)
          e^{i \omega t} P_{\ell} (\cos \theta)
                                               \\   \nonumber
\delta T^{(2)}_{r r}&=& \frac{1}{h}
\Bigg[ \left( \frac{3 h'}{2 r} + \frac{\lambda}{r^2} \right) H - 
        \frac{i}{\omega} \left\{ \frac{h'}{2} \left( \frac{h'}{r} + 
                                                    \frac{h + \lambda}{r^2}
                                                              \right)
                      - \frac{\omega^2}{r} \right\} H_1
                                                              \\  \nonumber
& &\hspace{4.0cm}
           + \left(\frac{h'^2}{4 h} - \frac{h'}{2 r} + \frac{\omega^2}{h}
                   - \frac{\lambda}{r^2} \right) K
                                                  \Bigg]
                     e^{i \omega t} P_{\ell} (\cos \theta).
\end{eqnarray}
Combining (\ref{defT3}) and (\ref{defT4}), one can derive a relation
\begin{equation}
\label{relation10}
\frac{i}{\omega} \left(h h' + \frac{r h'^2}{4} + r \omega^2 \right)
                                      \delta \Theta_{t r}
    + \left( 3 h h' + \frac{2 \lambda h}{r} \right) \delta \Theta_{r \theta}
    - \frac{i}{\omega} 
   \left[ \frac{(h + \lambda) h'}{r} + h'^2 - 2 \omega^2 \right]
    \delta \Theta_{t \theta} = 0.
\end{equation}
Thus one of the three non-diagonal components is not linearly independent.

The relation (\ref{relation10}) with the first-order differential equations
(\ref{first-order}) enables us to derive
\begin{equation}
\label{algebraic1}
\frac{d F}{d r} + \frac{h'}{2 h} F = \tilde{Q}(r)
\end{equation}
where
\begin{eqnarray}
\label{definetildeQ}
F&=&-(3 r h' + 2 \lambda) H + \left[2 i \omega r + \frac{1}{i \omega}
     \left\{ (h + \lambda) h' + r h'^2 \right\} \right] H_1
    + \left( 2 \lambda + r h' - \frac{r^2 h'^2}{2 h} - \frac{2 r^2}{h}
              \omega^2 \right) K
                                             \\   \nonumber
\tilde{Q}&=& - 2 \left(h' + \frac{r}{2} h'' \right)
            \left[ 3 H - \frac{i}{\omega} \left( 2 h' + \frac{h + \lambda}{r}
                                                 \right) H_1 + 
            \left(\frac{r h'}{h} - 1\right) K \right].
\end{eqnarray}
When $n=0$, $\tilde{Q}$ vanishes and, therefore, $F=0$ is used as an algebraic 
identity in Ref.\cite{zer70}. Thus Eq.(\ref{algebraic1}) is an corresponding 
algebraic identity for our case. 

Using $F$ in Eq.(\ref{definetildeQ}) one can eliminate $H$ by making use of  
\begin{eqnarray}
\label{elimi1}
H&=& \frac{1}{3 r h' + 2 \lambda}
\Bigg[ - F - \left(2 i \omega r + \frac{1}{i \omega}
      \left\{ (h + \lambda) h' + r h'^2 \right\} \right) H_1
                                                             \\  \nonumber
& &\hspace{4.0cm}
+ \left(2 \lambda + r h' - \frac{r^2 h'^2}{2 h} - \frac{2 r^2 \omega^2}{h}
                               \right) K \Bigg].
\end{eqnarray}
Then the first-order differential equations for $K$ and $H_1$ reduce to 
\begin{eqnarray}
\label{first-order2}
K'&=& \frac{i}{\omega} \delta \Theta_{t r} - \frac{F}{r(3 r h' + 2 \lambda)}
+\left[\alpha_0(r) + \omega^2 \alpha_2(r) \right] K + 
\left[ \beta_0(r) + \omega^2 \beta_2(r) \right] R
                                                    \\   \nonumber
R'&=& - \frac{2}{\omega h} \delta \Theta_{t \theta} + 
\frac{i F}{h (3 r h' + 2 \lambda)} + 
\left[ \gamma_0(r) + \omega^2 \gamma_2(r) \right] K +
\left[ \delta_0(r) + \omega^2 \delta_2(r) \right] R
\end{eqnarray}
where $R \equiv -H_1 / \omega$ and 
\begin{eqnarray}
\label{abc1}
& &\alpha_0 = \frac{h'(r h' - 2 h + \lambda)}{h (3 r h' + 2 \lambda)}
\hspace{4.0cm}
\alpha_2 = \frac{-2 r}{h (3 r h' + 2 \lambda)}
                                                       \\   \nonumber
& &\beta_0 = \frac{2 i}{r^2 (3 r h' + 2 \lambda)}
\left[r h' (r h' + h) + \lambda (\lambda + 2 r h' + h) \right]
\hspace{1.0cm}
\beta_2 = \frac{2 i}{3 r h' + 2 \lambda}
                                                      \\   \nonumber
& &\gamma_0 = -\frac{i (8 \lambda h + 8 r h h' - r^2 h'^2)}
                    {2 h^2 (3 r h' + 2 \lambda)}
\hspace{3.0cm}
\gamma_2 = \frac{2 i r^2}{h^2 (3 r h' + 2 \lambda)}
                                                       \\   \nonumber
& &\delta_0 = - \frac{h' (4 r h' + h + 3 \lambda)}{h (3 r h' + 2 \lambda)}
\hspace{4.0cm}
\delta_2 = \frac{2 r}{h (3 r h' + 2 \lambda)}.
\end{eqnarray}

Now, we perform the transformation
\begin{eqnarray}
\label{trans11}
K&=& f_1(r) \hat{K} + g_1(r) \hat{R}
                                        \\   \nonumber
R&=& f_2(r) \hat{K} + g_2(r) \hat{R}.
\end{eqnarray}
Then Eq.(\ref{first-order2}) reads
\begin{eqnarray}
\label{first-order3}
& &f_1 \hat{K}' + g_1 \hat{R}' = -\frac{i}{\omega} \delta \Theta_{t r} - 
\frac{F}{r (3 r h' + 2 \lambda)} + 
\left[-f_1' + (\alpha_0 + \omega^2 \alpha_2) f_1 + (\beta_0 + \omega^2 \beta_2)
                                                    f_2 \right] \hat{K}
                                                         \\  \nonumber
& &\hspace{5.0cm}
      + \left[ -g_1' + (\alpha_0 + \omega^2 \alpha_2) g_1 + 
                      (\beta_0 + \omega^2 \beta_2) g_2 \right] \hat{R}
                                                          \\  \nonumber
& &f_2 \hat{K}' + g_2 \hat{R}' = \frac{2}{\omega h} \delta \Theta_{t \theta}
        + \frac{i F}{h (3 r h' + 2 \lambda)} + 
    \left[-f_2' + (\gamma_0 + \omega^2 \gamma_2) f_1 + 
                                    (\delta_0 + \omega^2 \delta_2) f_2
                                                 \right] \hat{K}
                                                           \\   \nonumber
& &\hspace{5.0cm}
   + \left[-g_2' + (\gamma_0 + \omega^2 \gamma_2) g_1 + 
                     (\delta_0 + \omega^2 \delta_2) g_2 \right] \hat{R}.
\end{eqnarray}

As in the axial perturbation what can we do now 
is to choose two components of the 
energy-momentum tensor. For simplicity we choose as following
\begin{eqnarray}
\label{choice2}
& &\frac{i}{\omega} \delta \Theta_{t r} + 
\frac{F}{r (3 r h' + 2 \lambda)} = 0
                                       \\  \nonumber
& &\frac{2}{\omega h} \delta \Theta_{t \theta}
        + \frac{i F}{h (3 r h' + 2 \lambda)} = 0.
\end{eqnarray}
Since $\delta \Theta_{\mu \nu} = 0$ in $4d$ limit, Eq.(\ref{choice2})
implies $F = 0$ in the same limit. This is a boundary condition the 
authors in Ref.\cite{zer70,edel70} used for the derivation of the 
$4d$ effective potentials in the axial and polar perturbations. Thus 
Eq.(\ref{choice2}) guarantees the potential derived by making use of 
this equation has a correct $4d$ limit. Furthermore, Eq.(\ref{choice2})
makes Eq.(\ref{first-order3}) reduce to the simple expressions, which 
enables us to derive the effective potential analytically.

Then Eq.(\ref{relation10}) fixes the remaining non-diagonal component
$\delta \Theta_{r \theta}$. Expressing 
$f_1$, $f_2$ and $g_2$ in terms of $g_1$ as following
\begin{eqnarray}
\label{final1}
f_1&=& -h g_1' + \frac{3 r^2 h'^2 + 5 \lambda r h' + 2 \lambda h + 2 \lambda^2}
                      {r (3 r h' + 2 \lambda)} g_1
                                                        \\   \nonumber
f_2&=& i r \left[g_1' + \frac{3 r^2 h'^2 + 2 \lambda r h' - 4 \lambda h}
                             {2 r h (3 r h' + 2 \lambda)} g_1
                                                            \right]
                                                          \\   \nonumber
g_2&=& - \frac{i r}{h} g_1
\end{eqnarray}
and choosing $g_1$ as 
\begin{equation}
\label{final2}
g_1 = \left[ (n+1) \frac{r_H}{r^2 h'} \right]^{1 / 2 (n+1)}
      \left( \frac{2 \lambda r_H}{r (3 r h' + 2 \lambda)} 
                               \right)^{n / 2 (n+1)},
\end{equation}
one can make a Schr\"{o}dinger-like equation
\begin{equation}
\label{schro1}
\left( \frac{d^2}{d r_*^2} + \omega^2 \right) \hat{K} = 
V_P (r) \hat{K}
\end{equation}
where $r_*$ is defined as $d r / d r_* = h$ and 
\begin{eqnarray}
\label{final3}
& &V_P (r) = \frac{h}{r^2 (3 r h' + 2 \lambda)^2}
\Bigg[ 8 \lambda^3 + (24 r h' + 8 h + 2 r^2 h'') \lambda^2
                                                           \\  \nonumber
& & \hspace{5.0cm}
+ (18 r^2 h'^2 + 24 r h h' + 3 r^2 h h'' - 3 r^3 h h''') \lambda
                                                           \\  \nonumber
& & \hspace{5.0cm}
+ \left(18 r^2 h h'^2 + 9 r^3 h h' h'' + \frac{27}{4} r^4 h h''^2 - 
     \frac{9}{2} r^4 h'^2 h'' - \frac{9}{2} r^4 h h' h''' \right)
                                                          \Bigg].
\end{eqnarray}
When $h = 1 - (r_H / r)^{n+1}$, the effective potential (\ref{final3}) reduces 
to
\begin{eqnarray}
\label{final4}
& &V_P(r) = h \frac{r^{2 n}}{[2 \lambda r^{n + 1} + 3 (n + 1) r_H^{n+1}]^2}
                                                    \\  \nonumber
& & \hspace{3.0cm}
\times 
\Bigg[ 8 \lambda^3 + 2 \left\{4 - (n^2 - 9 n - 6) 
                               \left( \frac{r_H}{r} \right)^{n+1}
                                               \right\} \lambda^2
                                                     \\  \nonumber
& & \hspace{3.5cm}
-3 (n + 1) \left( \frac{r_H}{r} \right)^{n+1}
\left\{ n (n + 6) - (n^2 + 12 n + 6) \left( \frac{r_H}{r} \right)^{n+1}
                     \right\} \lambda
                                                       \\  \nonumber
& & \hspace{3.5cm}
+ \frac{9}{4} (n + 1)^2 \left( \frac{r_H}{r} \right)^{2n+2}
\left\{n (n - 2) + (n^2 + 8 n + 4) \left( \frac{r_H}{r} \right)^{n+1}
                                      \right\}
                                                   \Bigg].
\end{eqnarray}
It is easy to show that $V_P(r)$ has a correct $4d$ limit.
But same problem arises as in the case of the axial perturbation. 
The general principles we 
used cannot fix the energy-momentum tensor completely. Thus we choosed two
components in Eq.(\ref{choice2}). The different choice may yield a different
potential. Without knowing the exact nature of the matter on the brane it 
seems to be impossible to derive an unique effective potential. 

One may argue the situation may be improved if the Newman-Penrose formalism is 
applied. But there is also ambiguity in the Ricci terms which should be 
related to the energy-momentum tensor. Thus same problem occurs even in the 
Newman-Penrose formalism. 
%%There is another perturbation method called 
%%gauge invariant perturbation\cite{ger79}. However, it is shown
%%in Appendix that this method also does not improve the situation we 
%%encountered.

In this letter the polar perturbation is examined when the spacetime 
background is a metric induced on the brane from the bulk Schwarzschild
geometry. Since the projected background is not a vacuum solution of the 
Einstein equation, we tried to examine the behavior of the energy-momentum
tensor under the polar perturbation from the general principles. 
As in the case of the axial perturbation,
however, the general principles do not fix two components of the 
energy-momentum tensor. The suitable choice of these unfixed components
yields an effective potential which has a correct $4d$ limit. But the problem 
is the fact that different choice may yield a different potential. It is 
greatly nice if there is a physical mechanism which fixes all components
of the energy-momentum tensor completely.

\vspace{1cm}

{\bf Acknowledgement}:  
This work was supported by the Kyungnam University
Research Fund, 2006.

\newpage

\end{document}